\documentclass[12pt]{article}

\usepackage[english]{babel}

\makeatletter
\usepackage{latexsym}
\usepackage[T1]{fontenc}
\usepackage{amsmath}
\usepackage{amssymb}

\def \be {\begin{equation}}
\def \ee {\end{equation}}

\def \ume {{\scriptstyle{\frac{1}{2}}}}
\def \ra {\rightarrow}

\def \eqq {\equiv}

\def \b {{\beta}}

\def \eps {{\varepsilon}}

\def \ph {{\varphi}}

\def \psib {{\overline{\psi}}}

\def \A {{\cal A}}
\def \B {{\cal B}}
\def \C {{\cal C}}
\def \D {{\cal D}}

\def \G {{\cal G}}
\def \H {\mbox{${\cal H}$}}

\def \L {{\cal L}}

\def \O {{\cal O}}

\def \U {{\cal U}}
\def \V {{\cal V}}

\def \Z {{\cal Z}}

\def \hbf {{\bf h}}

\def \x {{\bf x}}

\def \Rbf {{\bf R}}

\def \Zbf {{\bf Z}}

\def \AO {{\cal A}({\cal O})}
\def \AO' {{\cal A}({\cal O}')}

\newfam\Ssfam
\font\eleSs=cmss10 at12pt \font\sevenSs= cmss10 at 8pt \font\sixSs= cmss10 at 6pt

\textfont\Ssfam=\eleSs\scriptfont\Ssfam=\sevenSs%
\scriptscriptfont\Ssfam=\sixSs \def\Ss{\fam\Ssfam\eleSs}

\def\doppio#1{{\rm I}\kern-.1667em{\rm #1}}

\def\Q{\text{Q}\kern-.52em
    \text{\vrule height1.5ex width.5pt depth0pt}\kern.45em}

\def \Zmath {{\mathchoice {\hbox{$\Ss\textstyle Z\kern-0.4em Z$}}
{\hbox{$\Ss\textstyle Z\kern-0.4em Z$}} {\hbox{$\Ss\scriptstyle Z\kern-0.25em
Z$}} {\hbox{$\Ss\scriptscriptstyle Z\kern-0.2em Z$}}}}

\def\Cmath{{\mathchoice{\hbox{$\rm\textstyle\text{\kern.35em\vrule
   height1.5ex width.5pt depth0pt\kern-.35em C}$}}
{\hbox{$\rm\textstyle\text{\kern.35em\vrule
   height1.5ex width.5pt depth0pt\kern-.35em C}$}}
{\hbox{$\rm\scriptstyle\text{\kern.35em\vrule
   height1.5ex width.3pt depth0pt\kern-.35em C}$}}
{\hbox{$\rm\scriptscriptstyle\text{\kern.35em\vrule
   height1.5ex width.2pt depth0pt\kern-.35em C}$}}}}

\def \be{\begin{equation} \displaystyle}
\def \ee{\end{equation}}

\def \A*{\mbox{$A^{*} $}}
\def \B*{\mbox{$B^{*} $}}
\def \C*{\mbox{$C^{*} $}}

\def \bea{\begin{eqnarray}}
\def \eea{\end{eqnarray}}

\def \b{\beta}

\def \bfsigma {\mbox{\boldmath${\sigma}$}}
\def \bftheta {\mbox{\boldmath${\theta}$}}

\@addtoreset{equation}{section}

\def \be {\begin{equation} \displaystyle}

\def \ee {\end{equation}}

\def \ra {\rightarrow}
\def\AO {\mbox{${\cal A}({\cal O})$}}
\def\AO'{\mbox{${\cal A}({\cal O}')$}}
\def\O {\mbox{${\cal O}$}}

\def\A{\mbox{${\cal A}$}}

\def \ra{\rightarrow}
\def \ph {{\varphi}}
\def \eps {{\varepsilon}}

\def \O {{\cal O}}

\def \A {{\cal A}}
\def \AO {\A(\O)}
\def \AOl'{\A(\O_{loc}')}

\def \B {{\cal B}}

\def \D {{\cal D}}

\def \H {{\cal H}}

\def \x {{\bf x}}

\begin{document}
\begin{titlepage}

\title{The strong CP problem revisited and solved  by  \\the gauge group topology}

\sloppy

\author{F. Strocchi  \\  Dipartimento di Fisica, Università di Pisa, Pisa, Italy}

\fussy
\date{}

\maketitle

\begin{abstract}

We exploit the non-perturbative result that the $\theta$ angle which defines  the vacuum structure is not a $c$-number free parameter, as suggested by the instanton semi-classical approximation, but instead one of the points of the spectrum of the central operator $\theta_{op}$ which describes the gauge group topology. Hence, the value of such an angle should not be \textit{a priori} fixed, but rather be determined, as any quantum operator, by the infinite volume limit of the functional integral, where the fermionic mass term uniquely fixes the phase, with $<\theta_{op} > = \theta_M$, the mass angle. Such an equality is stable under radiative corrections performed before the infinite volume limit of the functional integral; this provides a solution of the strong $CP$ problem.
The mechanisms is carefully controlled  in the massive Schwinger model with attention to the infrared problems, to the volume effects induced by boundary terms and with a careful discussion of the infinite volume limit of the functional integral.
 In the QCD case, a chirally symmetric finite volume functional measure may be also obtained by 
modifying the  Atiyah, Patodi and Singer spectral boundary conditions, which play a crucial role in the standard approach; such modified  boundary conditions give a  fermion determinant which  is real and therefore chirally  and CP symmetric.

\end{abstract}

\noindent {\bf{Mathematical Subject Classification}}: 81R40,     81T13,   81V05

\noindent {\bf{Keywords}}:  Quantum Chromodynamics, Gauge group topology, \\ Strong $CP$ problem
\end{titlepage} 

\section{Introduction}
The strong CP problem has been raised as a consequence of the addition of the topological term 
\be{\L_\theta = i\, \theta \,(64 \pi^2)^{-1}\, \mbox{Tr}\,[\tilde{F} F]}\ee  to  the QCD Lagrangian $\L^{QCD}$. 

The standard wisdom relies on the instanton semi-classical approximation and on the  topological classification   of the   instanton solutions by the winding number $n$ of their \textit{large distance behavior} (assumed to be a pure gauge). 

According to this point of view, the functional integral is written as the sum over $n$ of the integrals over euclidean field configurations with winding number $n$ (\cite{WE}, Section 23.6)
\be{ Z = \sum_n e^{i \theta n } \int_n  D A_\mu  D \psi D \bar{\psi  }\, e^{- \int d x\, \L^{QCD}_n}= \int  D A_\mu  D \psi D \bar{\psi  }\, e^{- \int d x \, (\L^{QCD} +\L_\theta)}. }\ee
In such an approach, crucially conditioned by the instanton semi-classical approximation, the  $\theta$ angle  is a ($c-$ number) \textit{free parameter}. 

The topological term has been argued  to play the role of a \textit{symmetry breaking boundary term} which, in the infinite volume limit of the functional integral, selects the representation of the observables given by a corresponding $\theta$ vacuum (\cite{WE}, see also the discussion in Section 4 below) and then  leads to chiral symmetry  breaking.

The topological term is $CP$ odd and leads to a strong $CP$ breaking unless $theta$  coincides with the  chiral angle $\theta_M = \mbox{Arg det} M$  of the quark mass  $M$. Even if with a fine tuning such an equality   is imposed at the tree level, it  is not stable under radiative corrections; on the other side,  the lack of detection of  the neutron electric dipole moment $d_n$ indicates that   the effective angle 
$\theta_{eff} \eqq \theta -  \theta_M$ must be less than 10$^{-9}$.  The explanation of such a small bound in a natural way is the so-called strong $CP$ problem.

\vspace{1mm} 
As long as the instanton semi-classical approximation has played a unique role for establishing the $\theta$ vacuum  structure, with chiral symmetry breaking,   the standard strategy has been the only available recipe for the analysis of the functional integral and of the vacuum structure; however, the realization of a  deeper explanation of the vacuum topological properties (\cite{MS09}; \cite{FS21}, Appendix F; \cite{FS19}),  opens a new perspective and an alternative approach to the strong $CP$ problem. The gauge group topology has been shown to provide a solution of the $U(1)$ problem and we shall argue that it also lead to a solution of the strong $CP$ problem. 

We shall revisit the  analysis of the strong $CP$ problem,
 with a non-perturbative approach which exploits  the following  established results \cite{FS21, FS19} (a brief account in Appendix A below):

\def \thop { \theta_{op}}
\vspace{1mm}
\noindent  i) the proof that \textit {the vacuum structure is provided by the spectrum of the  unitary operators} $T_n = \exp{ i\,n \thop}$, with $n \in \Zbf$, the winding number which classifies of the   gauge transformations; the topological operators $T_n$  belong to the center of the local algebra of observables and  their spectrum, ($\{e^{i\, n\theta}$, $\theta \in [0, \,2\pi)\}$), which is not pointwise invariant under chiral symmetry,  labels the irreducible (more generally factorial) representations of the observables, each with  a corresponding $\theta$ vacuum and chiral symmetry breaking. Thus, the $\theta$ label does not appear as a free parameter  to be  \textit{a priori} fixed, but its value,   corresponding to the  ground state expectation of $\thop$,  should rather be  determined by the infinite volume limit of the functional integral (as any other operator of the theory).

The  crucial role of the \textit{gauge group topology},  represented by the central operators $\exp{ i\,n \thop}$, 
 makes clear the limits of the standard strategy, conditioned by the instanton semi-classical approximation, since the \textit{a priori} fixed $\theta$  precludes the effect of the fermion mass term in (uniquely)  determining the expectation  of $\thop$; as we shall argue below, the introduction  of the mass term, which plays a role similar to that of an external field, does not commute with the infinite volume limit even if it is a small perturbation (a standard result in statistical mechanics, see also Section 2 below); 

\vspace{1mm}
\noindent ii) the existence of the chiral anomaly does not imply chiral symmetry breaking, as often stated in the literature, but only the fact that the gauge invariant chiral current  $j^5_\mu$ does not generate \textit{the chiral transformations} of the fields (and in particular of the observables) which   are actually \textit{well defined} and  generated  by the (formal) exponentials of the  conserved gauge dependent chiral current $J^5_\mu$; as we shall discuss below (Sections 3, 4), with chirally symmetric  boundary conditions (leading to a chirally symmetric finite volume functional measure) the emergence of chiral symmetry breaking is  devolved to the infinite volume limit of the functional integral, to be taken \textit{after} the introduction of the quark mass term; 

\vspace{1mm}
\noindent iii) 
the possible dangerous mismatch between the  topological term and $\theta_M$  is ruled out  by the infinite volume limit with chirally symmetric boundary conditions, since    in this case the value of the operator $\thop$ is not \textit{a priori } forced  by  symmetry breaking  boundary conditions, but it is rather  driven by the quark mass term  to coincide  with $\theta_M$ in the  \textit{unique} infinite volume limit of the euclidean functional integral, minimizing the free energy; 
 the infinite volume limit of the functional integral, with chirally symmetric   boundary conditions, treats the topological term and the mass term on equal footing  for the emergence of chiral symmetry breaking, which in the standard approach  is
 \textit{ a priori} decided  by the choice of the $c$-number free parameter $\theta$, with the  mass term playing   a secondary role; 
\goodbreak

\vspace{1mm}
\noindent  iv)   it is usually stated in the literature that boundary terms (as the topological term) do not have any effect on the dynamics and that  their role is essentially confined  to the  determination of the ground or equilibrium state,  but the tacit underlying assumption is the short range of the dynamics; in fact, as discussed in Section 2 (see \cite{FS21} for a general account), in the presence of long range interactions \textit{boundary terms may give rise to volume effects} which  affect the equations of motion and compete with an external field.

\vspace{1mm}

In Section 3 the above  mechanism for the solution of the strong $CP$ is explicitly  controlled in the massive Schwinger model, which shares  the structural  features of QCD believed to be the crucial ingredients for the breaking of chiral  symmetry and $CP$ symmetry: chiral anomaly, chiral transformations of the fields, euclidean solutions with non-trivial winding number, topological term, mass term with a chiral angle, $CP$ problem etc..

 A special care in taking the infinite volume limit  and in handling the boundary terms leads to significantly different conclusions with respect to previous analyses  of the massive Schwinger model which are  obtained by freely integrating by parts, neglecting the corresponding boundary terms  \cite{CJS, CO1, CO2}. In particular, the infinite volume limit of the functional integral with free boundary conditions taken after the introduction of the mass term yields a solution of the  $CP$ problem (the standard strategy and the one advocated here lead to an equivalent physical picture if there is no mass term).  

In our opinion,  the non-perturbative realization that the $\theta$ angle which labels  the vacuum is  given by (the spectrum of) the unitary operator which describes the gauge group topology, represents a radical improvement  with respect to the instanton semi-classical approximation, (which relies on regularity properties which fail even in the free field case). Moreover, it allows to use, as in the Schwinger model, boundary conditions  which modify the Atiyah, Patodi, Singer (APS) boundary conditions and do not break chiral symmetry.  

If there is no fermionic mass term,
one gets a chirally symmetric mixed phase with chiral breaking arising in each pure phase as in the standard approach. However, in the presence of a mass term, in the infinite volume limit one gets a unique phase  defined by a vacuum state labeled by an angle which coincides with the mass angle $\theta_M$ (for an explicit discussion of such a mechanism see  Sections 3, 4). 




\newpage
\section{Boundary terms and volume effects}

According to the standard wisdom, boundary terms have no effect on the dynamics and their role is to select the states, typically the ground state (in the infinite volume limit). Indeed, rigorous results on statistical mechanics \cite{RU}, as well as explicit controls of models confirm this picture, but in general it is not sufficiently emphasized that (essential) locality of the interaction is the underlying crucial assumption. 

Indeed, in the case of short range interaction the coupling of a variable $A$, say localized at the origin, with variables localized on the boundary of a sphere $V(R)$ of radius $R$, vanishes in the infinite volume limit and therefore boundary terms $B$ in the Hamiltonian do not affect the dynamics of localized variables. On the other side, if  the interaction has a long range decrease, say in three dimensions as $R^{-2}$, the interaction with the boundary variables does not vanish in the infinite volume limit and actually \textit{variables at infinity} may enter in the time evolution (\cite{MS87}, \cite {FS21} Appendix A).

A prototype of such variables at infinity is the infinite volume limit of averages of variables on   the boundary 
\be{  \lim_{R \ra \infty} B_{\partial V(R) } \eqq \lim_{R \ra \infty} \int d x f_R(x)  B_x ,}\ee
where  $\partial V$ denotes the boundary of $V$, $B_x $  the $x$- translated of $B$,
 and $f_R
$ is a regular function, with supp $f_R = \{ R < |x| < R(1 + \eps)\}$, and $\int d x f_R(x) = 1$.
\footnote{ The existence of the (weak) limit is guaranteed if  the states to which it is applied satisfy a  regular large distance  behavior; such infrared regularity condition is satisfied by a large class of phases 
defined by a translationally invariant ground (or equilibrium)   state (for a more detailed mathematical discussion see \cite{MS87}). Asymptotic abelianess is taken for granted, in order to assure a reasonable quantum mechanical interpretation:
\be{ \lim_{|x| \ra \infty}  [ \,A,  \,B_x\,] = 0, \,\,\,\,\,\,\forall A, B \in \A.}\ee } 

Such variables at infinity are localized outside any bounded region and coincide with the corresponding ergodic means
\be{ B_\infty = \lim_{V \ra \infty} \frac{1}{V} \int_V \, d x\, B_x. }\ee
Thus,  as a consequence of a long range interaction,  a boundary term gives rise  to a volume effect leading to  the appearance of a variable at infinity in the time evolution of local variables. 

By asymptotic abelianess, the variables at infinity  commute with all the essentially localized variables and therefore  belong to the center $\Z$ of the algebra $\A$ generated by them  (\textit{central variables}).  

\def \bfsigmtext   {{\boldmath$\sigma$}}
 \def \bfsigma {\mbox{\boldmath${\sigma}$}}
\def \bftheta {\mbox{\boldmath${\theta}$}}

    This mechanism is displayed by the Heisenberg model with long range interaction. For simplicity and for making the message more direct, we shall discuss  the molecular field approximation described by the following (finite volume) Hamiltonian involving the coupling with the boundary variable $\bfsigma_{\partial V}$ : 
\be{ H_V    = - J \sum_{i  \in V} \bfsigma_i \cdot \bfsigma_{\partial V},  \,\,\,\,\,\,\,\bfsigma_{\partial V} \eqq \frac{1}{|\partial V|} \sum_{j \in \partial V} \bfsigma_j.       }\ee
In the infinite volume limit $ \bfsigma_{\partial V}$ converges  to $\bfsigma_\infty$ (in the   large class of infrared regular states) giving rise to a volume effect similar to a coupling with an external field. 

If the infinite volume limit is taken with free boundary conditions one gets a mixed phase,    
 with $\bfsigma_\infty$ acting as an operator in the correspondent representation space, where the symmetry of spin rotations is therefore unbroken; however,   the decomposition   according to the spectrum of $\bfsigma_\infty$ yields pure phases  labeled by the  expectation $\bftheta \eqq < \bfsigma_\infty >_0 $ on the correspondent pure ground (or equilibrium) state; 
the spin rotations are broken in each pure phase leaving unbroken the rotations  around  $\bftheta$.

In this simple model, the interaction with the boundary variable has infinite range, but the same conclusion holds with a boundary interaction of the form
 $$ - \sum_{i  \in V,\, j \in \partial V} J_{i \, j}\, \bfsigma_i \cdot \bfsigma_j,$$ 
with $J_{i \, j }$ of long range, say  $\sim |i - j |^{-2}$ in three dimensions, since in any representation defined by a translationally invariant state, satisfying asymptotic abelianess, with the property that the center $\Z$  is invariant under translations, one has  (see e.g. \cite{FS21}, Proposition 16.3): 
\be{ \lim_{|x| \ra \infty} A_x  = A_\infty.}\ee

Thus, in the case of long range  interactions, the volume effect induced by a boundary term may  compete with the effect of an external field interaction, say $ \hbf \cdot \sum_{i \in V} \bfsigma_i$. 
In fact, in the infinite volume limit with free boundary conditions  the equations of motion are 
\be{ i \,\frac{d \bfsigma_i}{ d t}  = 2 (\hbf  - J \bfsigma_\infty) \wedge \bfsigma_i;}\ee
the presence of the external field removes the ground state degeneracy and selects the unique ground (or equilibrium) state on which the central  variable  $\bfsigma_\infty$ gets aligned to the external field $\hbf$:
\be{ 0 = i\, \frac{ d < \bfsigma_i >_0}{d  t} = 2(\hbf - J < \bfsigma_\infty >_0) \wedge  < \bfsigma_\infty >_0.}\ee
The symmetry breaking is induced only by $\hbf$, so that  
the rotations around $\hbf$ are not broken. 

As an alternative, one may destroy the operator character of the boundary term $\bfsigma_{\partial V}$ by freezing it to one of the points $\bftheta$ of the spectrum of $ \bfsigma_\infty$, which labels the pure phases in the absence of the external field $\hbf$.   In the infinite volume limit, with such an enforced boundary condition, where $\bftheta$ plays the role of a free parameter, the finite volume Hamiltonian takes the form
 \be{H_V =   \sum_{i \in V} \bfsigma_i \cdot ( \hbf - J \, \bftheta).}\ee
Then, the residual symmetry is the group of rotations around  $\hbf - J \bftheta$, the alignment  of $< \bfsigma >_0$ with $\hbf$  is not "natural"  and to this purpose a "fine tuning" is needed. 

In the absence of the external field $\hbf$, the two strategies lead to the same final picture, with pure phases labeled by the order parameter $\bftheta$ (in the first case  the direct sum of them, in the second case the unique phase selected by the chosen boundary value $\theta$). But there is a substantial  difference if $\hbf \neq 0$; as a  consequence of   the long range interaction with the boundary, \textit{the infinite volume limit and the introduction of the external field do not commute}, even if the interaction with the external field can be treated as a small perturbation. 

The lesson of the model is that in the presence of an external field the choice of free boundary conditions leads to a residual symmetry determined by the external field with  no disturbing effect by the boundary term.
 
It is worthwhile to note that it is the first strategy with free boundary conditions which gives the correct (infinite volume extrapolation of the) behavior of a spin system with long range coupling to the boundary in the presence of a homogeneous external magnetic field, which  forces the spin on the boundary to get aligned to it (for minimizing the free energy).  

The general picture displayed by the above simple model applies also to the Heisenberg model with long range coupling to the boundary.

As we shall see below a very similar structure  characterizes the strong CP problem, where the quark mass matrix plays the role of an external field driving the value/expectation of the topological operator $\tilde{\theta}$ to a value coinciding with  the mass angle.

\sloppy
\section{A lesson from the Schwinger model. \\  Boundary terms and CP symmetry}
\fussy

As a useful laboratory for exporting the above strategy to QCD, we consider the Schwinger model in the axial  gauge in the bosonized version, with a special care in taking the infinite volume limit  and in handling the boundary terms. 
The discussion shall be done in the Lagrangian formulation
in view of the functional integral approach to quantum fields.

To this purpose a warning is at stage for theories with long range interactions. 


The standard use of the Lagrangian is the derivation of the Euler-Lagrange (EL) equations by the stationarity of the action $A$ under local variations of the fields, but in order to get information on the boundary effects on the dynamics one needs additional information beyond the local EL equations; this is obtained by  exploring the  stationarity of the action under variations which feel the space boundary. This  has not been  considered in the literature by the underlying implicit assumption of a local dynamics. 

To this purpose we exploit the  stationarity of the action in an arbitrary  space time volume $V$ (typically $V = V_S \times  [-T, \,T]$,  $V_S$ a sphere centered at the origin) under arbitrary infinitesimal  variations of the fields $\delta_V  \ph(x, t) = \eps \chi(x, t)_V $ with $\chi_V$ infinitely differentiable, and vanishing at the time boundary,  $\chi_V(x, -T) = 0 = \chi_V(x, T)$:

$$ 0= \delta A _V = \int_{x \in V}  d^3 x\, d t\, \eps \chi_V \left [ \frac{\delta \L}{\delta \ph} - \partial_\mu \frac{\delta \L }{\delta \partial_\mu \ph} \right ]  +$$ 
\be{+ \int d t \,\eps \chi_V(x, t) \, n_i(x, t)) \frac{\delta \L}{ \delta \partial_i \ph(x, t)} \mid_{x \in \partial V_S},}\ee
where $n_i(x, t)$ is the unit vector normal to the space boundary, $\partial V_S$. The first term on the right hand side vanishes by the stationarity of the action under variations which vanish on the space boundary and yield the EL equations.  

The last term is the result of the integration by parts and automatically  vanishes if the fields are localized inside $V_S$, but otherwise it  gives information on  the behavior of the fields on the space boundary ({\bf{\textit{Boundary equations}}}), since the variations $\chi_V(x, t)$ are allowed to take arbitrary value on $V_S$: 
\be{n_i(x, t) \frac{\delta \L}{ \delta \partial_i \ph(x, t)} \mid_{x \in \partial V_S } \,=\,0.}\ee

These equations  provide the needed additional information for a full control of the dynamics including the boundary effects (see below). 

In particular,  a four divergence $\partial K \eqq \partial ^\mu K_\mu$ gives a non-vanishing contribution  to the boundary equations
and it is responsible for boundary terms in the time evolution of the fields. For a more detailed discussion of this delicate point,  usually overlooked in the literature,  see \cite{MS90}.

We start by considering the massless Schwinger model. The euclidean Lagrangian is 
\be{\L  = - \ume [ (\partial _0 \phi)^2 + (\partial_1 \phi)^2 + (\partial_1 A_0)^2 ] + i (e/\sqrt{\pi}) A_0 \partial_1 \phi   + i (e/2\pi) \theta \,\partial_1 A_0,}\ee 
where the last term is the topological term (with the free parameter $\theta$).

\def \limL   {\lim_{L \ra \infty}} 
\def \thetati  {\tilde{\theta}}
\def  \tie  {\tilde{e}}

\def \thetabar   {\theta/2\sqrt{\pi}}
\def \ebar   {e/2 \sqrt{\pi}}

In order to correctly treat the infrared effects and the boundary terms we put the system on a lattice of finite size $V = (2 L + 1) \times (2 L +1)$,  so that we can work with free or periodic boundary conditions, which do not break chiral symmetry.

The stationarity of the action integral in the space-time volume  $V_S \times T$, $T = [t_1, t_2]$, with respect to infinitesimal variations of the fields $ \delta \phi(x, t) = \eps \chi(x, t) =\delta A_0(x, t)$, with  $\chi(x, t_1) = 0 = \chi(x, t_2)$,  
gives the following euclidean equations ($ \tie \eqq e/ \sqrt{\pi}$) 
\be{\Delta \phi - i \tie \,\partial_1 A_0 = 0, \,\,\,\,\,\,\,\,\partial_1 [ i \tie\, \phi + \partial_1 A_0 ] = 0.}\ee
The second equation corresponds to the local Gauss law; the equation may be integrated   yielding    $ [ i \tie \phi + \partial_1 A_0 ] =  c, \,\partial_1 c = 0$ and the integration constant $c$, is determined by the Boundary equations (BE):
\be{[ - \partial_1 \phi - i \tie A_0 ] |_{x \in V_S} = 0, \,\,\,\,\,\,\,\,\,\,[ - \partial_1 A_0 + i \tie\, \thetati ] |_{x \in V_S}  = 0 \,\,\,\,\, \Longrightarrow\,\,\,\,\,\,\,c = i \tie (\phi_B + \thetati),}\ee 
where  $\thetati \eqq \theta/ 2 \sqrt{\pi}$ and  $\phi_B$ denotes the value of $\phi$ on the space boundary.  Then, one gets the following equations 
\be{\Delta\, \phi_{i \,j} = \tie^2 (\phi_{i\, j} - \phi_{B\,j} - \thetati), }\ee
\be{ E_{i\, j} = i \tie   (\phi_{i\, j} - \phi_{B\,j} - \thetati),}\ee
where $\Delta$ is the discrete Laplacian, $  \phi_{i\, j}  $ is the lattice version of $\phi$, $(i\, j)$ denote the lattice site indices in $[ -L, L] \times [ -L, L]$, $ \phi_{B\,j} \eqq \ume ( \phi_{L\, j} + \phi_{-L\, j})$ and $E = - \partial_1 A_0$ is the electric field.

A delicate issue is the infinite volume limit, which requires a careful handling of the infrared structure. In the correlation functions of $\phi$ (more precisely of $\exp{2 i \sqrt{\pi} \phi}$) and $E$ one obtains\footnote{ By an analysis of the spectrum of the operators which define the action as a quadratic form \cite{MS90} and Appendix B below .}
\be{ \limL \phi_{B\,i} = \phi_\infty  - \thetati,    }\ee
with the central variable   $$ \phi_\infty \eqq \lim_{V \ra \infty}   V^{- 1} \sum_{i, j \in V} \phi_{i j}, $$
the  limit being defined on the infinite volume correlation functions of $E$ and $\exp{2 i \sqrt{\pi} \phi}$. Then, one gets the following equations 
\be{\Delta\, \phi = \tie^2\, (\phi - \phi_\infty), \,\,\,\,\,\,\,\,\,\, E = i \,\tie\,   (\phi - \phi_\infty) ,}\ee
invariant under the chiral symmetry $\phi \ra \phi + \alpha/\sqrt{\pi}$.

 Thus, $ \phi = \phi_\infty + \phi_0,$ with $\phi_0$ a free field of mass $\tie^2$ and vanishing mean value (since $< \phi > = < \phi_\infty >$).

For displaying the crucial role of the boundary terms and their interplay with the topological term it is convenient to single out the part $A_l$ of the action which gives rise to euclidean equation without boundary terms, (namely corresponding to the choice of  $\phi_B = 0, A_0|_B =0$, which breaks chiral symmetry):
$$\Delta \phi = i\,\tie \,\partial_1 A_0, \,\,\,\,\,\,\,E = - i\,\tie \phi. $$   
Hence, $$A_l = \int d^2\,x\, \{ - \ume\,[ (\partial_0 \phi)^2 + (\partial_1 \phi)^2 ] - \ume (\partial_1 A_0)^2 - i \,\tie\,\partial_1 A_0 \, \phi \}, $$ (which is not chirally invariant) and 
$$ A = A_l + i \,\tie\, \int d^2 x \,\partial_1 (A_0\,\phi)  + i\,\tie\, \thetati \int d^2 x \,E. $$
In the infinite volume limit, the value of $\phi$ on the boundary (in the second term on the right hand side) may be replaced by $\phi_B $,  which converges to $\phi_\infty - \thetati$ in such a limit. Then,
$$ A = A_l + i\,\tie \,\phi_B \int d^2 x  \,E +  i \tie \,\thetati  \int d^2 x \,E \sim $$
\be {\sim A_l + i\, \tie \,\phi_\infty \int d^2 x\, E.} \ee 
Thus, by the effect of the infrared and boundary terms,  in particular by eq.\,(3.8), \textit{the  $c-$number $\theta$ angle } in the topological terms gets \textit{replaced by the (central variable) operator $\phi_\infty$}. 

This result agrees with the non-perturbative analysis of chiral symmetry breaking in QCD, \cite{FS21} (Appendix F), \cite{FS19},  according to which the 
$c-$number $\theta$ angle, introduced on the basis of the instanton semiclassical approximation \cite{WE}, should be rather replaced by the central variable $\theta_{op}$ of the  topological operator. In fact, the operator $T \eqq \exp{ i (\phi - E/\tie)}$ plays the role of the generator of the "large" gauge transformations and on the physical states describes the topology of the gauge group; furthermore,  in Minkowski space the second of eqs.\,(3.9) becomes $E = \tie\,(\phi - \phi_\infty)$ and $ T = \exp{ i\, \phi_\infty}$. 

Equation (3.10) establishes a strong connection with the spin models discussed in Section 2, and, as discussed in that case, two strategies are available both giving an essentially equivalent  picture (in the massless case). 

 In fact, if one uses free boundary conditions in the infinite volume limit one gets a chirally symmetric mixed phase, since the action (3.10) is chirally invariant; then, the  decomposition according to the spectrum of $\phi_\infty$ yields pure phases labeled by  an angle $\theta = < \phi_\infty >$, with the emergence of  chiral symmetry breaking in each of them. 

On the other side, one may use symmetry breaking boundary conditions, with $\phi_\infty$  replaced  by a ($c$-number) angle  $\theta$ in eq.\, (3.10); since by  a chiral transformation  $A_l$ changes by a term equivalent to a shift of $\theta$,  the so obtained  action $A = A_l + i \, \tie \thetati \int d^2 x \, E$ is no longer
chiral invariant and in  this way one directly gets   the pure phase labeled  by the symmetry breaking order parameter $ < \phi_\infty > = \theta$.

As in the case of the spin models discussed above, the two strategies do no longer lead to  equivalent results if one adds an external field, namely in the presence of a fermion mass term $ M :\cos(2 \sqrt{\pi} \phi - \theta_M):$ , where $: \,\,: $ denotes the Wick ordering and   $\theta_M$  is the chiral angle of the mass term.  

The second strategy gives rise to a mismatch between the topological term and the chiral angle of the fermion mass; in a certain sense, this corresponds to first treating the chiral symmetry breaking in QCD and afterwards the effect of the electroweak interaction,  running  against the unity of the 
Standard Model where strong and electroweak interaction coexist from the start. 

As discussed in Section 2, the fermion mass term may significantly  affect/influence the value taken by the  (central) operator $\phi_\infty$ in the    
topological term which, according to eq.\,(3.10),  replaces  the $c-$number angle $\theta$. 

 \def \thetatim {{\thetati_M}}
In order to see the radical difference of the two strategies we shall consider  the massive Schwinger model with the mass term $M  :\cos (2 \sqrt{\pi} \phi - \theta_M):$ replaced by $ m^2 :(\phi - \thetatim)^2:$\,, $\thetatim \eqq \theta_M/ 2 \sqrt{\pi}$. 

Instead of eq.\, (3.6) one has  
\be{\Delta\, \phi_{i \,j} = \tie^2 (\phi_{i\, j} - \phi_{B\,j} - \thetati)  + m^2( \phi_{i \,j} - \thetatim), }\ee whereas eq.\,(3.7) remains unchanged.
 
The standard strategy corresponds to \textit{a priori} fixing the $\theta$ angle, equivalently to putting $\phi_\infty = \theta $ in eq.\,(3.10); by eqs.\,(3.8),  this is obtained by chosing the boundary condition $\ph_B = 0$, which \textit{a priori} breaks chiral symmetry, independently from the mass term.  

Then the vacuum expectation of eq.\,(3.11) gives  
$$< \phi > = (\tie^2 \,\thetati + m^2 \,\thetatim)/(\tie^2 + m^2)$$ and 
\be{ < E > = i\, \frac{m^2\, \tie\, (\thetatim - \thetati)}{ \tie^2 + m^2}.       }\ee
Hence, if $\theta \neq \theta_M$ one has $< E > \neq 0$ and $CP$ symmetry  is violated.

On the other hand,  if the value of $\phi_\infty$ in eq.\,(3.8) is not \textit{a priori} fixed,  equivalently  by letting  $< \phi >$ to be determined by the functional integral in the \textit{unique} phase selected by the mass term, one can prove, \cite{MS90} and Appendix A below,  that in that phase  $\phi_\infty$ can only take the value $\thetatim$ and therefore, by eq.\,(3.8), eqs.\,(3.11, \,3.7)  become 
\be {\Delta \phi = (\tie^2 + m^2) (\phi - \thetatim),}\ee
\be {E = i \,\tie (\phi - \thetatim).}\ee
Thus, since $ < \phi > = < \phi_\infty >$, one has 
\be{ <  E  > = i \,\tie \, (< \phi  > - \thetatim) = 0; }\ee
there is no background electric field and $CP$ symmetry is not  broken.  
  \vspace{1mm}

The lesson by the model is the following:
\vspace{1mm}

\noindent 1) since the vacuum $\theta$ angle is not a $c$-number free parameter, but the value of an operator (which describes the gauge group topology) \cite{FS21}, it should not be fixed \textit{a priori} but let it  be determined by the infinite volume limit of the functional integral. In fact, by eq.\,(3.10) the action is chirally symmetric and so is the functional integral with free boundary condition; this  leads to a sort  of degeneracy described by   the spectrum of the topological operator $\phi_\infty$ and, in the infinite volume limit, the fermion mass term removes such a degeneracy by fixing $<\phi_\infty > = \thetatim$

\vspace{1mm} 
\noindent 2) the infinite volume limit with periodic or free boundary conditions, which do not break chiral symmetry, avoids a mismatch between the topological  term and the angle of the  mass term  and yields a solution of  the $CP$ problem.
\goodbreak 



\section{Gauge group topology and a solution of the strong $CP$ problem}
According to the lesson of the spin models discussed in  Section 2 and of the massive Schwinger model, for the solution of the strong CP problem  one should avoid  to \textit{a priori} break chiral symmetry with a non-symmetric boundary term and rather let  the mass terms  select the  unique phase in the infinite volume limit.
   To this purpose we discuss the chiral transformations of the finite volume functional measure.  

\def \Mth   {M_{\theta_M}}
\def \DsAM   { \not\!\D_A + M}
\def \DsAMt  {\not\!\D_{A} + M_{\theta_M}}
\def  \detDsAM    {\mbox{det}(\not\!\D_{A} + M)}
\def \psib   { \bar{\psi}}
\def \psipsib  {\psi_1...\psib_n}
\def \DsA  {\not\!\D_A}
\def \detDsAMt  {\mbox{det}(\not\!\D_{A} + M_{\theta_M})}

Formally, the correlation functions in finite volume $V$ are  given by 
\be{ < A_1 ...\psi_j ...A_k...\psib_n >^V_{M_{\theta_M}}   = \int d\mu_V(A)\, A_1...A_k... {\det}( \not\!\D_A + \Mth)_V < \psi_j...\psib_n >^V_A, }\ee
 where $d\mu_V(A)$ is the measure over the gauge fields $A_i$,
 \be{ d\mu_V(A) = \prod_{x \in V} d\,A_x \exp\left ( \int_V  d x \, \L_A \right ), }\ee
\be{  < \psi_j...\psib_n >^V_A = Z^{-1}\int D\psi\, D\psib\, \exp \left [ \int d^4 x \,\L_{\psi A} \right]  \psi_j ...  \psib_n},\ee
are the fermionic  correlation functions, with $D \psi, D\psib$ the Berezin fermonic integration, with the normalization corresponding to the fermionic  correlation functions in an external gauge field $A$, $Z = \det( \not\!\D_A + \Mth)$, 
\be{ \not\!\D_A \eqq \gamma_\mu (\partial_\mu + i g A_\mu ), \,\,\,\,\,\,\,\,
 \Mth \eqq m (\cos \theta_M + i \gamma_5 \sin \theta_M),}\ee
 $ \L_{\psi A}$ is the fermionic part of the Lagrangian and  $\L_A$ the purely gauge part.
\vspace{2mm}

\noindent {\bf {\textit{ a)  Ultraviolet regularization of the fermionic correlation functions}}}

\vspace{1mm}
It is sometimes stated in the literature that the ultraviolet (UV) regularization of the fermionic correlation functions, in particular for the definition of the chiral current, is responsible for the breaking of chiral symmetry. As shown in \cite{FS21} this is not correct, since the chiral transformations of the fields (in particular of the observables) are well defined and represented by the  unitary (formal) exponentials   of the conserved (gauge dependent) chiral current $J^5_\mu$ . 

The structure is that of a spontaneous symmetry breaking, without Goldstone bosons because the crucial assumption of the Goldstone theorem fails, namely the vacuum expectations of the  infinitesimal chiral transformations of gauge invariant operators are not given by the  expectations of the commutator with a local current \cite{FS21}. 

The chiral anomaly does not imply the lack of chiral symmetry of the functional measure in finite volume, which  only depends on the boundary conditions, as we shall discuss below.

We  use boundary conditions which  define $\DsA$ as an anti-hermitian operator ($\gamma_\mu^\dagger= \gamma_\mu$) and map the domain of  $\DsAM$ into itself. Since $M$ is hermitian  it precludes the existence of zero modes for $\detDsAM$.
Then, one may  prove \cite{MS90} that the UV regularization which cuts the high eigenvalues of $\DsAM$ and the regularization with the standard UV cutoff in the interaction, 
give the same fermionic correlation functions.  

The independence of the fermionic correlation functions (in an external gauge field) from the UV regularization does not imply   that the  compound fields are  independent from the UV regularization adopted for their definition; for example, the gauge invariant regularization which cuts the large eingenvalues of $\DsA$ yields the  anomalous gauge invariant chiral current, $j^5_\mu$,  whereas the chiral invariant UV regularization  which cuts the large eigenvalues of the free Dirac operator leads to a chiral current $J^5_\mu$,  which is conserved in the limit of massless fermions, but  it is not gauge invariant.

Thus, the possible lack of chiral symmetry of the functional measure in finite volume, being decided by the correlation functions only depends on the boundary conditions, not on the UV regularization adopted in the construction of the gauge invariant  anomalous chiral current, as often stated in the literature. This is clearly displayed by the Schwinger model discussed before.         
\vspace{2mm}

\def \Mth   {M_{\theta_M}}
\def \DsAM   { \not\!\D_A + M}
\def \DsAMt  {\not\!\D_{A} + M_{\theta_M}}
\def  \detDsAM    {\mbox{det}(\not\!\D_{A} + M)}
\def \psib   { \bar{\psi}}
\def \psipsib  {\psi_1...\psib_n}
\def \DsA  {\not\!\D_A}
\def \detDsAMt  {\mbox{det}(\not\!\D_{A} + M_{\theta_M})}



\noindent {\bf {\textit{ b)  Boundary term and chiral symmetry  of the functional \\ measure}}}

\def \bal {\beta^\alpha}
\def \DsA  {\not\!\D_A}

\def \DsA'  {\not\!\D_{A^'}}

\def \DsAM   {\not\!\D_A + M}
\def \DsAMz  {\not\!\D_A + M_0}
\def \DsAMt   {\not\!\D_A + M_{\theta_M}}

For $M = 0$, the chiral symmetry of the functional measure (in finite volume) means the invariance of the correlation functions under the following  chiral transformations, $\b^\alpha, \alpha \in \Rbf,$ of the fields 
\be{ \beta^\alpha (\psi) = e^{i \alpha \gamma_5}\,\psi, \,\,\,\,\,\,\bal(A_\mu) = \,A_\mu.}\ee

For $M \neq 0$, by chiral symmetry of the functional measure we  mean the invariance of the correlation functions under the above chiral transformations (4.5) accompanied by  
\be{\theta_M   \ra  \theta_M + 2 \alpha.}\ee 

Now,  one has $ \DsAMt = e^{i \theta_M \gamma_5/2} (\DsAMz) \, e^{i \theta_M \gamma_5/2} $, ($M_0 \eqq M_{\theta_M = 0}$), which implies 
\be{(\DsAMt)^{-1} = e^{- i \theta_M \gamma_5/2} (\DsAMz)^{-1} \, e^{-i \theta_M \gamma_5/2}. }\ee
Thus, since  by Berezin integration the fermionic correlation functions are obtained in terms of  finite numbers of eigenvalues and eigenfunctions   of\,\,\,\,\,\,
$(\DsAMt)^{-1}$, they  transform covariantly under (4.5), (4.6).

The control of the contribution of $\det(\DsAMt)_V$ to the functional measure in finite volume, eq.\,(4.1), requires a regularization. We shall choose a regularizaton which cuts the large eingenvalues  in a symmetric way, namely $ |\lambda_n| \leq \Lambda$, then we have 
\be{\det(\DsAMt)  = \prod_{ |\lambda_n| \leq \Lambda, \lambda_{0, n} \neq 0} [\det(\DsAMt)]_{\lambda_n}  [\det(\DsAMt)]_0,}\ee
where the subscript $\lambda_n$ denotes the restriction to the block corresponding to the pairs $\lambda_n, \bar{\lambda}_n$, such that $\lambda_{0, n} \eqq \lim_{M \ra 0} \lambda_n \neq 0  $   and the subscript $0$ the restriction to the space generated by the
 $n_{\pm}$ modes, with $\lambda_{0, n} = 0$; then
\be{  [\det(\DsAMt)]_{\lambda_n}  = | \lambda_n|^2,  \,\,\,\,\, [\det(\DsAMt)]_0 = e^{ i\, \theta_M (n_+ - n_-)}\, m^{n_+ +n_-},}\ee
so that $\det(\DsAMt) $ is not chiral invariant if $ n_+ - n_- \neq 0$. 



The index theorem gives the following relation
\be{n_+ - n_- = \nu \eqq - (1/64 \pi^2) \mbox{Tr} \int d x  \tilde{F}\,F, }\ee
apart from corrections which become irrelevant in the infinite volume limit, so that the functional measure in finite volume is not chirally invariant in the presence of gauge fields with $\nu \neq 0$, as indicated by the existence of instantons.
It should be stressed that the index theorem holds if the so-called Atiyah, Patodi and Singer (APS) spectral boundary conditions are imposed; in fact, the dependence of $\nu$ on $A_\mu$ only arises through the dependence of the boundary conditions on $A_\mu$.

Since under a chiral transformation of the fermionic fields  $\theta_M \ra \theta_M + 2 \alpha$, the effect is equivalent  to shifting the  $\theta$ angle, $\theta \ra \theta + 2\alpha$,  of the topological term in the standard approach, so that the $\theta$ term is not chirally symmetric  and so is also the vacuum selected by it in the infinite volume limit.

In the standard approach \cite{WE, CO}, the form of the topological term with the appearance of the $\theta$ angle is dictated by the requirement that the  correlation functions of QCD satisfy the cluster property. Indeed, if QCD is considered as isolated from the rest of the Standard Model (SM),  such a requirement is unavoidable for a reasonable physical interpretation of the theory.
 \goodbreak
 However, this is no longer necessary if one consider the Standard Model in its unity with QCD a part of it, since the validity of the cluster property may be guaranteed by other terms of the  SM, in particular by the fermionic mass term.

\def \thop {\theta_{op}}

Hence, we advocate the replacement of the $c$-number $\theta$ angle  in the topological term by the topological central operator $\thop$, whose spectrum has been proved to label the vacuum representation. 

\noindent This choice has the following properties

\vspace{1mm}
\noindent i) avoids to \textit{a priori} break  chiral symmetry of the finite volume functional measure with a symmetry breaking boundary term (of the standard approach);

\vspace{1mm}
\noindent ii) lets the fermionic mass term play a primary role in the chiral symmetry breaking by choosing the unique pure phase labeled by a point of the spectrum of $\thop$, which cannot be anything else that $\theta_M$, $< \thop > = \theta_M$, since the selected pure phase  has the same symmetry of the external field  (see Sections 2, 3);

\vspace{1mm}
\noindent iii) in the pure QCD theory (with no fermionic mass term) yields a chirally symmetric mixed phase, with a decomposition into pure phases according to the spectrum of $\thop$, with a final physical picture equivalent to that of the standard approach;

\vspace{1mm}
\noindent iv) it agrees with the lesson of the Schwinger model, see eq.\,( 3.10), which displays the replacement of the $c$-number $\theta$ angle of 
the topological term by the  central (topological) operator $\phi_{\infty}$;

\vspace{1mm}
\noindent v) leads to a solution of the strong $CP$ problem which afflicts  the standard approach.

\vspace{1mm}

The relevant point is that in the case of short range interactions the role of symmetry breaking boundary conditions or of boundary terms is to select the pure phases of an otherwise symmetric theory, with spontaneous symmetry breaking; but in the presence of a non-symmetric external field the boundary terms become irrelevant.

A substantial difference characterizes the case of long interactions so that the presence of a non-symmetric boundary term must be treated with much care, since  
it gives rise to volume effects mismatching the effect of an external field.

In the QCD case, the appearance of the  topological central  operator $\thop$, rather than of the free $c$-number $\theta$, better complies with the non-perturbative analysis, by which  the topological vacuum structure is better accounted for  by the spectrum of $\thop$, rather than by the  $\theta$ angle arising in  the instanton semi-classical approximation.

Then,  one is led to replace the  finite volume functional integral of the standard approach, eq.\,(1.2), by   
 \be{Z_{sym} =
 \int  D A_\mu  D \psi D \bar{\psi}\, e^{- \int d x \, (\L^{QCD}  + \L_{\thop})},\,\,\,
 \,\,\,\,\L_{\thop} \eqq - i\, \thop \,\frac{1}{64 \pi^2}\, \mbox{Tr}\,[\tilde{F} F]. }\ee

Hence, in presence of the fermionic mass term, the correlation functions in finite volume are now given  by

$$ \hspace{-70mm} < A_1 ...\psi_j ...A_k...\psib_n >^V_{M_{\theta_M}}   = $$
\be{= \int d\mu_V(A)\, A_1...A_k... {\det}( \not\!\D_A + \Mth)_V e^{-i \L_{\thop}} < \psi_j...\psib_n >^V_A, }\ee
 where $d\mu_V(A)$ and 
$< \psi_j...\psib_n >^V_A$
are defined as  in eqs.\,(4.2, 4.3).

By eqs.\,(4.9, 4.10), under a chiral transformation $\theta_M \ra \theta_M + 2 \alpha$ and
\be{{\det}( \not\!\D_A + \Mth)_V   \,\,\ra\,\,{\det}( \not\!\D_A + \Mth)_V \,
e^{i \,2 \alpha\, \mbox{Tr} [\tilde{F} F]/(64 \pi^2)};}\ee
moreover, the chiral transformation of the unitary topological operators $T^n$ (see Appendix A), imply that $\thop \,\,\ra \thop - 2 \alpha$. Then, the finite volume functional measure is chirally symmetric as in the Schwinger model (eq.\,(3.10)).

The replacement of $\theta$ by $\thop$ essentially amounts to incoherently sum the vacuum representations labeled by the $\theta$ angle; the functional integral of eq.\,(4.12) corresponds to integrate the last term of eq.\, (1.2) with the spectral measure $d \mu(\theta)$ of the topological operator $T = e^{i\,\thop} = \int d \mu(\theta) \, e^{ i\, \theta} $, i.e. to incoherently sum/integrate over the incoherent phases identified by the points of the spectrum of $\thop$. In the absence of the mass term, this gives a chirally symmetric mixed phase represented by a Hilbert space $\H = \int d \mu(\theta) \H_\theta$.

According to the strategy advocated in this note, one should not dissect the chiral symmetry breaking, by first treating its breaking in QCD and by then introducing  the effect of the electroweak interaction;  one should rather use a chirally symmetric  action (in finite volume) for the whole Standard Model, and devolve the symmetry breaking to the Higgs vacuum expectation (likely a fermionic condensation), which gives rise to the masses  and governs the breaking of chiral symmetry  in its unitary wholeness.

\vspace{2mm}
\noindent {\bf {\textit{ c) CP symmetry}}}
\vspace{1mm}
\def \Mt  {$M_{\theta_M}$}
\def \npm {$n_+ = n_-$}


If $CP_M$ and $CP_0$ denote the $CP$ symmetries defined by $M_{\theta_M}$  and by $M_{\theta = 0}$,  one has
\be{(\DsAMt)^{CP_M} = e^{i \theta_M \, \gamma_5/2}\,(\DsAMz)^{CP_0}\,e^{i \theta_M \, \gamma_5/2}}\ee
and the spectrum of $(\DsAMz)^{CP_0}$ is obtained from that of $\DsAMz$ by replacing each eigenvalue $\lambda_n$ by its complex conjugate  $\bar{\lambda}_n$. Then, the $CP_M$ symmetry of the fermionic correlation functions in an external gauge field follows from their chiral covariance (for $\gamma_5$ invariant boundary conditions). 

\noindent As far as the functional measure over the gauge field configurations is concerned,  for $\theta_M = 0$ the $CP_M$ symmetry follows from eqs.\,(4.8), (4.9), for $\gamma_5$ invariant boundary conditions. 

For $\theta_M \neq 0$,   by eq.\,(4.9), the spectrum of $\DsAMt$ is not invariant under complex conjugation, $\lambda_n \ra \bar{\lambda}_n$,  the determinant is not real  (with a complex phase dependent on $\theta_M$) and therefore not $CP_M$ symmetric. 

The complex phase  of ${\det}( \not\!\D_A + \Mth)_V$ may be eliminated by a redefinition of the fermionic fields corresponding to  a chiral transformation with parameter $\alpha = -  \theta_M/2$. However, as a result of this chiral transformation $\thop \,\ra \, \thop - \theta_M$ and the topological term becomes 
$\exp\{- i \,(\thop - \theta_M) \,\mbox{Tr} \,[ \tilde{F}\,F]/(64 \pi^2)\}$.
Since $\mbox{Tr} \,[ \tilde{F}\,F]$ is $CP$ odd, this term breaks $CP$ symmetry.

In the standard approach, eq.\,(1.2), (with the $c$-number $\theta$ angle), $CP$ symmetry requires $\theta= \theta_M$ to all orders of perturbative renormalization; since $\theta = \theta_M$ at the tree level is not stable under radiative corrections, a (not natural) order by order fine tuning would be necessary  to  get strong $CP$ symmetry.  

On the other side, with the strategy advocated above, in the infinite volume limit (performed at the very end), the fermionic mass term  selects the unique pure phase with the same symmetry of the fermionic mass and therefore  labeled by the point of the spectrum of $\thop$  which cannot be anything else than equal to $\theta_M$,  $< \thop > = \theta_M$.
Thus, the pure phase selected by the mass term (acting as an external field)  is described by  a  finite volume functional  measure with  a real ${\det}( \not\!\D_A + \Mth)_V$ and with  a   topological term which disappears in the infinite volume limit  (\textit{solution  of the strong $CP$ problem}).


\vspace{1mm}
\noindent {\bf{\textit{d)  Modified APS boundary conditions}}}
\vspace{1mm}

Correlation  functions corresponding to $\theta = \theta_M$ may also be obtained  by using modified Atiyah, Patodi, Singer  (MAPS) spectral boundary conditions \cite{MS90}, which yield a real ${\det}( \not\!\D_A + \Mth)_V$ and lead to a  chirally  symmetric functional measure in finite volume, (without losing gauge invariance \cite{MS90}) with  no $CP$ violating topological term.
 
The APS  spectral boundary conditions play a crucial role for the derivation of the index theorem, which in turn leads to a complex 
(and therefore $CP$ non-invariant) ${\det}( \not\!\D_A + \Mth)_V$, see eqs.\,(4.9, 4.10). 

After the proposal  \cite{MS90}, the MAPS boundary conditions have been further analyzed from a mathematical point of view, leading to a work-stream  in the general   theory of elliptic  boundary problems for Dirac operators \cite{BMSW, WSMB, BB}. They share the following properties of the  APS boundary conditions,


\def \DsA  {\not\!\D_A}

\noindent 1) antiself-adjointness of $\not\!\D_A$,

\noindent 2) discrete spectrum for $\not\!\D_A$ with finite multiplicity and having only $\lambda = \infty$ as accumulation point,

\noindent 3) $\gamma_5$ invariant domain of  $\DsA$, 

\noindent 4) gauge invariance of $\not\!\D_A$, namely if $A'$ is  related to $A$ by a globally defined gauge transformation $\U$, then   $\not\!\D_{A'} = \U\,\not\!\D_A\,\U^{-1}$,

\noindent 5) $CP$ invariance of the domain of $\not\!\D_A$ 

\vspace{1mm}
\noindent But, in contrast with  the APS case, they give

\noindent 6)  $n_+- n_-$ = independent of $A_\mu$.

\vspace{1mm}
For the detailed definition and proof of 1)-6) see \cite{MS90} and Appendix B below. It is worthwhile to remark that the APS boundary conditions are singled out if, in addition to the gauge invariant condition 4), one further requires that if $A_\mu$ is a pure gauge around the boundary, i.e. $A_B = \V \,\partial \,\V^{-1}$, with 
\textit{ $\V$ not
 globally defined on $V$ if $\nu \neq 0$} (a rather strong condition from a physical point of view), then the corresponding boundary condition is obtained from that for $A = 0$ by acting with $\V$. On the other hand, for the modified APS boundary conditions
 the gauge invariance constraint on the boundary  is required only if two configurations of $A_\mu$ are related by a \textit{gauge transformation over the whole volume} $V$.

Since for $M \neq 0$,  $A = 0$ and $\gamma_5$ invariant boundary conditions,  there are no zero modes, in the following we shall take  $n_+ = n_-$.

 As a consequence of  eq.\, (4.7), the fermionic correlation functions in external gauge field transform covariantly under chiral transformations. Moreover, with the choice $n_+ = n_-$, by eqs.\,(4.8), (4.9), $\det(\DsAMt)_V$ is independent of  $\theta_M$ and therefore chirally  invariant. Hence, its fermion loop expansion does not require counter terms of the form of the topological term, with  a non-zero $\theta$. 

The resulting (finite volume) functional measure on $A$ is therefore invariant under chiral transformations and the correlation functions  (4.1) transform covariantly  under chiral transformations
\be{< A_1 ...\psi_j ...A_k...\psib_n >^V_{M_{\theta_M}} = < A_1 ...e^{ -i \,\theta_M \gamma_5/2}    \psi_j ...A_k...\psib_n\, e^{ -i \,\theta_M \gamma_5/2} >^V_{M_{\theta_M= 0}}.}\ee




This mechanism is further displayed by a comparison with  the standard approach. 
Since  the modified APS boundary conditions yield \npm and $|\lambda_n|^2$ is independent of $\theta_M$, eqs.\,(4.9), (4.10) (for large $V$) give
 \be{(\det (\DsAMt))^{MAPS} = \det (\DsAMz)  \simeq [\det (\DsAMt)\, e^{- i \theta_M (n_+ - n_)}]^{APS}, }\ee
where the superscript MAPS denotes the use of modified APS (advocated here)  and  APS denotes the determinant and its phase obtained in the standard approach  by using APS boundary conditions. 

\noindent Furthermore, in the limit $V \ra \infty$ the operator  $\DsAM$ is unique and so is $(\DsAM)^{-1}$ which defines the fermionic correlation functions in an external gauge field; therefore they become independent of the boundary conditions and are the same as those of the standard approach. \goodbreak

\noindent Then, in the limit of  large $V$, one has  
\vspace{1mm} 
$$ \hspace{-2.8 in} < A_1 ...\psi_j ...A_k...\psib_n >^{MAPS}_{M_{\theta_M}}   \simeq$$ 
$$ \int d\mu(A)\,[\det (\DsAMt) \, e^{- i \theta_M (n_+ - n_-)})]^{APS}\,
 A_1...A_k...  < \psi_j...\psib_n >_{A, {M_{\theta_M}}} = $$
$$= \int d\mu_V(A)
([{\det}( \not\!\D_A + \Mth) e^{- i \theta (n_+ - n_-)}]^{APS}
  A_1...A_k...  < \psi_j...\psib_n >_{A, {M_{\theta_M}}})_{\theta = \theta_M} $$ 
\be{\,\,\,\,\,\,\,\,\,\,\,\,\,\,\,\,\,\,\,\,\,\,\,\,\;\;	\;	\;\;\;\; = (< A_1 ...\psi_j ...A_k...\psib_n >^{APS}_{M_{\theta_M}})^{\theta = \theta_M}.}\ee
The correlation functions on the right hand side are those obtained according to the standard wisdom with $\theta = \theta_M$. However, the perturbative renomalization of the two theories is very different.

In the standard approach, if in the presence of the mass term  one puts $\theta = 0$ in the topological term, one has $CP_M$ symmetry at the tree level , but by eqs.\,(4.8),(4.9), $\det(\DsAMt)$ is complex, so that its fermion loop expansion requires counter terms of the form of the topological term, i.e. a non-zero $\theta$. In order to keep $CP_M$ symmetry one needs a fine tuning  $\theta =\theta_M$ order by order in the fermion loop expansion (even if $\theta_M$ is not renomalized by the strong interactions). In fact, after taking the sum over fermion loops, the contribution to the effective action 
\be{{\det}( \not\!\D_A + \Mth) e^{- i \theta (n_+ - n_-)}]^{APS}_{\theta = \theta_M}
}\ee
is independent of $\theta_M$, real and $CP_M$ symmetric. However, such a restoration of $CP$ symmetry through a fine tuning is not stable under the addition of the weak interactions, since if one puts  $\theta = \theta_M^{tree}$ (where the superscript $tree $ means the value at the tree level in the weak interactions) one does not get a $CP_M$ symmetric expansion in the weak coupling constant $g_W$.

 On the other side, with the choice of chirally symmetric boundary conditions, as e.g. the modified APS boundary conditions, ${\det}( \not\!\D_A + \Mth)$ is real and independent of $\theta_M$ and therefore $CP_M$ symmetric at each order of the weak expansion of $\theta_M$.

 The point is that in the standard approach, the equality $\theta = \theta_M$ is not assured by a symmetry, since the finite volume functional measure is not chirally symmetric, whereas by  the choice of chirally symmetric boundary conditions, the vanishing of $CP$ violating parameter is guaranteed by the chiral symmetry (in the sense of eqs.\,(4.5), (4.6)), of the finite volume functional measure. 


\section{Appendix A: Local gauge group topology and its unitary representation}
As discussed in \cite{FS21}, the local gauge group is parametrized by $C^\infty$ unitary  functions $\U$ with value in the global gauge group $G$, differing from the identity only in   compact sets  in space and at most polynomial growth in time;  in the temporal gauge  the residual local  gauge group $\G$ is defined by time independent   gauge functions  $\U$  and it is represented by unitary operators $V(\U)$, \cite{FS21}. 

The Gauss subgroup $\G_0 \subset \G$ is defined by unitary gauge functions continuously connected with the identity $\U(\lambda g), \lambda \in \Rbf, \,g(\x) = g_a(\x) T^a, \,\,g_a$ of compact support in space, (in the following briefly denoted by $\U_g$). 
The subspace of physical state vectors $\Psi$  is identified by the subsidiary condition of invariance under  $\G_0$: $V(\U_g) \Psi = \Psi, \,\,\, \forall \U_g \in \G_0$; we denote by $P_0$ the projection on such a subspace. 

The gauge functions $\U$ fall into disjoint homotopy classes labeled by the winding number $n(\U)$ and the topological group is the group $\G/\G_0$. It is an abelian group represented by unitary operators $T_n \eqq P_0\, V(\U_n) \,P_0$ on the subspace of physical states. 
The $T_n$ are local gauge invariant unitary operators and therefore belong to the center of the (local) observable algebra $\A$. The points of its spectrum $\{e^{i n \theta}, \theta \in [0, 2 \pi) \}$ label the irreducible (more generally factorial) representations of $\A$, providing the non-perturbative  topological explanation of the $\theta$ vacua.  

The chiral transformations $\beta_5^\lambda, \, \lambda \in \Rbf$, are well defined on the field algebra and locally implemented by   unitary operators $V^5_R(\lambda)$, \textit {formally} the exponentials $e^{ i \lambda J^5_0(f_R \alpha)}$, ($J^5_\mu$ the conserved chiral current, $J^5_0(f_R \alpha) = \int d^4 x\, J^5_0(x) f_R(\x) \alpha(x_0)$, \,
 $f_R(\x) = 1$ for  $|\x| \leq R$, $f_R(\x) = 0$ for $|\x| \geq  R(1 + \eps)$, $\alpha(x_0)$ of compact support, with $\int d x_0 \,\alpha(x_0) = 1$,  \cite{FS21}).
The non-trivial topology of $\G$ implies that 

\noindent i) $\beta_5^\lambda(T_n) = e^{- 2 i n \lambda} T_n$ , so that the chiral symmetry is \textit{always} broken in any irreducible/factorial representation of $\A$;

\noindent ii) the local implementers of the chiral transformations $V^5_R(\lambda)$ are not weakly continuous in the group parameter $\lambda$, so that (in the vacuum expectations) the infinitesimal chiral transformations are not generated  by the commutator with a local current, a crucial assumption of the existence    of Goldstone bosons \cite{FS21}, (\textit{solution of the $U(1)$ problem}).

 The above topological structure only  relies on the differentiability of  the gauge functions $\U$, which is part of the definition of  $\G$, whereas in the standard approach  the vacuum structure is derived by assuming regularity properties  of the instanton configurations, which give them zero functional measure  (e.g. \cite{CO}  Appendix C, \cite{FS16} Chapter  5, Section 8).


\section{Appendix B:  Boundary terms in the Schwinger model}   
We control the infinite volume limit of $\phi$ and $\phi_B$, in the continuum limit in the temporal gauge. The equations (3.11), (3.7), with $\phi_{i j}, \,E_{i j }$, replaced by $\phi,  E = -\partial_0 A_1$ and $\phi_B \eqq \ume [\,\phi(x_1, T) + \phi(x_1, -T)\,]$  are now supplemented by the boundary equations on the time boundary $\partial T$ and on the space boundary $\partial V_S$:
\be{ E = -i  \,\tie\,\thetati, \,\,\mbox{on}\,\,\,\partial T,\,\,\,\,\,\,\,\, \,\,
       \partial_0 \phi  = i \tie A_1, \,\,\mbox{on}\,\,\,\partial T, \,\,\,\,\,\,\,\,\,\,\partial_1 \phi = 0, \,\,\,\mbox{on}\,\,\,\partial V_S. }\ee 
By completely fixing the gauge with the condition $A_1(x_1, T) = - A_1(x_1, - T)$, the second of the boundary conditions  eqs.\,(6.1) takes the form
\be{\partial_0 \phi(x_1, \pm T) \mp \ume \tie^2 \int_{-T}^T d x'_0\, (\phi - \phi_B)  = \mp \tie^2 \,T\,\thetati,}\ee  
and, by eq.\,(3.7), the first of eqs.\,6.1) is equivalent to $\phi(x_1, T) = \phi(x_1, - T)$.

We first consider the case $\theta_M = 0$.
  
\noindent With such boundary conditions, eq.\,(3.11) has a unique solution of class $C^2$ in the interior of $V_S \times [ - T, T]$ and of class $C^1$ on the boundary. Such  a solution can be found by Fourier expansion  $\phi(x_1, x_0)= \sum_n c_n(x_0)\,\phi_n(x_1)$, in terms of eigenfunctions of $\partial_1^2 $, $\phi_n(x_1) = \cos(\pi n x_1/T)$. 

The general solution is $c_n = 0,$ for $n \neq 0$,
\be{c_0(x_0)  =  a \cosh(\omega_0 \,x_0)  + b_0^1,}\ee
 where $\omega_0 \eqq (\tie^2 + m^2)^{\ume}$, $b_0^1$ is a particular solution  of the inhomogeneous equation: $b_0^1 = (\thetati + c_B) \tie^2/(\tie^2 + m^2)$. Moreover, the constants $a, c_B$ are determined by the condition $c_B = c_0(T)$ and by the boundary condition on $\partial_0 c_0(\pm T)$:
\be{ a = - \thetati\,(\cosh(\omega_0T))^{-1} \frac{\tie^2 \omega_0 T}{ \tie^2 \omega_0 T + m^2 \tanh(\omega_0 T)}, }\ee
\be{ \thetati + c_B = - \thetati (\tie^2 + m^2)  \frac{\tanh(\omega_0 T)}{ \tie^2 \omega_0 T + m^2 \tanh(\omega_0 T)}. }\ee

The solution in the case $\theta_M \neq 0$ is obtained by a chiral transformation, i.e. by a shift: $c_0(x_0)  \ra c_0(x_0) + \thetatim$. 

The equations (6.3)-(6.5) imply that in the limit $T \ra \infty$
\be{ \lim_{T \ra \infty} < \phi >_T = \thetatim, }\ee
everywhere, except for a region of size $1/\omega_0$.

Furthermore, one has 
\be{ < \phi_B >_T = - \thetati  + \thetatim + O(1/\omega_0 T).}\ee

\section{Appendix C: Modified APS boundary \\conditions}

We briefly recall the APS boundary conditions. 

In finite volume $V$, one considers gauge fields which are pure gauge around the boundary of $V$ and one  writes the euclidean Dirac operator  in a neighbourhood of  the (closed) boundary, defined by $r = 0$,  in the form $i \gamma_r\,(\partial_r +B(A))$, where the subscript $r$ denotes the component in the direction normal to the boundary. 
Furthermore, one can assume that $A_\mu $ has no component normal to the boundary. 

Since $B_A$ is a differential operator defined on a compact manifold, its spectrum is discrete and one may denote by    $H^A_{<0}$, and $H^A_{\geq 0}$, the subspaces of states generated  by eigenstates  of the restrictions  $B(A)_+$, $B(A)_-$, of $B(A)$, respectively   to positive, respectively negative, chirality states. 

Then, the \textit{APS boundary conditions} require:
\be{ \psi|_{r=  0} \in H^A_{< 0}, \,\,\,\,\,\,\mbox{if} \,\gamma_5 \psi = \psi; \,\,\,\,\,\,  
 \psi|_{r=  0} \in H^A_{\geq 0}, \,\,\,\,\,\,\mbox{if} \,\gamma_5 \psi = -\psi.}\ee
The above definition  implies that if on the boundary the field configurations $A$ and $A = 0$ are connected by the gauge transformation $U(A)$, i.e. $$A_\mu|_{r = 0} = U(A) \partial_\mu U(A)^{-1}|_{r = 0},$$
 then  $H^A_{< 0} = U(A) \,H^{A = 0}_{< 0}|_{r = 0}$ and similarly  for 
$H^A_{\geq 0}$. 

The \textit{modified APS boundary conditions} are defined in the following way: for each sector $ [\nu] $ of gauge configurations with topological number $\nu(A) = \nu$,  one chooses a representative configuration $A^{(\nu)}$ and requires 
$$ \psi|_{r=  0} \in K ^A_{< 0} \eqq U(A)\,U^{-1}_{\nu(A)} \,H^{A = 0}_{< 0}, \,\,\,\,\,\,\,\mbox{if}\,\, \gamma_5 \psi = \psi,$$
 where $U^{-1}_{\nu(A)}$ is the gauge transformation which on the boundary connects $A = 0$ to the fixed  gauge configuration $A^\nu$.

 Similarly, one defines $K^A_{\geq 0}$ and requires 
$\psi|_{r = 0} \in K^A_{\geq 0}, \,\,\,\,\,\mbox{ if }  \,\,\,\,\,\,\gamma_5 \psi = -\psi.$

Thus, the modified boundary conditions for the reference gauge configurations $A^\nu$  are the APS boundary  conditions for the free Dirac operator ($A = 0$), since 
$$K^{A^\nu}_{< 0} = H^{A = 0}_{< 0}, \,\,\,\,\,\,K^{A^\nu}_{\geq 0} = H^{A = 0}_{\geq 0}.$$
\goodbreak

\section{Conclusions}

In the standard approach the strong $CP$ problem is due to the non-vanishing  coefficient $(\theta - \theta_M)$ in the topological term  (its vanishing at the tree level is not stable under perturbative renormalization).

The strategy suggested in this note is to replace  the $c$-number $\theta$ angle (which in the standard approach plays the role of labeling the vacuum states)  by the central topological operator $\thop$, which has been proved to better describe the non-perturbative topological vacuum structure.

Such a modification does not substantially  change the physical picture in the absence of the fermion mass term, the basic starting point in the standard approach, but it represents a drastic  change  in the presence of the fermion mass term providing a solution of the strong $CP$ problem; in fact, according to the general wisdom of Statistical Mechanics, in the infinite volume limit the mass term selects the unique pure phase labelled  by a point $\theta$ of the spectrum of $\thop$, which coincides with $\theta_M$; $\theta = < \thop > = \theta_M$. Thus, after elimination of the phase of the fermion determninant, the coefficient of the topological term vanishes in the infinite volume limit (performed at the very end).

From a general point of view, the idea is that QCD should not be considered as isolated from the rest of he Standard Model (SM); in particular, one should not first discuss chiral symmetry breaking in pure QCD and only then  confront it with the chiral breaking of the electroweak interaction. In fact, the introduction of the $\theta$ boundary term, with  $\theta $  a $c-$ number free parameter,   \textit{a priori} breaks chiral symmetry irrespectively of the role of the mass term.

The replacement of $\theta$ by $\thop$, which in the infinite volume limit is crucially sensible to the effect of the mass term, reaffirms the unity of the Standard Model.

\vspace{10mm}


\end{document}